\documentstyle[psfig]{caosp}
\begin{document}
\def\teff{$T_{\rm eff}$}
\def\lgg{$\log\,{g}$}
\def\vt{$\xi_{\rm t}$}
\def\vsini{$v\cdot \sin i$}
\def\kms{\,km\,s$^{-1}$}
\def\and{$\alpha$ And}
%%%%%%%%%%%%%%%%%%%%%%%%%%%%%%%%%%%%%%%%%%%%%%%%%%%%%%%%%%%%%%%%%%%%%%%%%%%%%%
\pubyear{1998}
\volume{27}
\firstpage{356}
%%%%%%%%%%%%%%%%%%%%%%%%%%%%%%%%%%%%%%%%%%%%%%%%%%%%%%%%%%%%%%%%%%%%%%%%%%%%%%
\htitle{Orbital elements and elemental abundances of $\alpha$ And}
\hauthor{T. Ryabchikova {\it et al.}}
\title{The double-lined spectroscopic binary $\alpha$ Andromedae: 
        orbital elements and elemental abundances }
\author{T. Ryabchikova \inst{1}, V. Malanushenko \inst{2}
and S. J. Adelman \inst{3}}
\institute{Institute of Astronomy, RAS, Moscow, Russia
\and Crimean Astrophysical Observatory, Nauchny, Crimea, Ukraine
\and Department of Physics, The Citadel, 171 Moultrie Street, Charleston,\\ 
SC 29409, USA}
\maketitle

\begin{abstract}
We performed a spectroscopic study of the SB2 Mercury-Manganese star
$\alpha$ And. Our measurements of the secondary's radial velocities result in
improved orbital elements. The secondary shows abundances typical of the
metallic-line stars: a Ca deficiency, small overabundances of the iron-peak
elements, and 1.0 dex overabundances of Sr and Ba.
\end{abstract}

\section{Introduction}

\vspace{-1mm}
\and\, (HD\, 358, HR\, 15) is a bright, well known binary system with a HgMn
primary. Pan et al. (1992) obtained its visual orbit from observations with the
Mark III Stellar Interferometer. Later Tomkin et al. (1995) detected the
weak secondary spectrum, measured the secondary's radial velocities near the
nodes, and calculated the spectroscopic orbital elements of this SB2 system.
Their derived mass ratio 0.42 $\pm$ 0.02 differs from 0.48 of Pan et al. and
gives too large a mass for the primary. No abundances of the secondary are
published.

\vspace{-3mm}
\section{Observations and orbital elements}

\vspace{-1mm}
\and\, was observed in 1990-91 and 1996-97 at Crimean Astrophysical
Observatory (CrAO) and in 1992-94 at Dominion Astrophysical Observatory (DAO).
CrAO spectra were obtained at the Coud\'{e} spectrograph of the 2.6m telescope
with a CCD detector in three spectral regions centred at
$\lambda\lambda$ 4960, 6347 and 6678 with S/N $\geq$ 300.
The DAO Reticon spectra were taken with the 1.22m telescope for a
grid of central wavelengths between $\lambda\lambda$ 3830 and 5180 with
55~\AA\, offsets.  The typical S/N was $\geq$ 200.

Radial velocity measurements for both components were made using 
synthetic spectra calculated with the model atmospheres and abundances 
discussed below.  For each spectral region we synthesized the primary's spectra
and derived the radial velocities from the shifts of the calculated spectrum
relative to the observations.  The rotational velocities are \vsini=52 \kms 
for the primary and \vsini=110 \kms for the secondary.

We calculated the orbital elements for \and  ~~with a code by Tokovinin (1992)
combining our measurements with those from Abt \& Snowden (1973), Aikman (1976),
and Tomkin et al. (1995). Table 1 compares these orbital elements.

\begin{table}
\caption{Orbital elements of \and}
\label{table1}
\begin{flushleft}
\begin{small}
\begin{tabular}{lccc}
\noalign{\smallskip}
               & Aikman(1976)        & Tomkin et al.(1995)& Present           \\
P(days)        &96.6960 $\pm$ 0.0013 &96.6963 (fixed)     &96.7051$\pm$0.0030 \\
T(JD2400000+)  &42056.32$\pm$ 0.20   &49212.17$\pm$ 0.20  &48245.49$\pm$ 0.23 \\
e              &0.521$\pm$ 0.008     &0.60 $\pm$ 0.02     &0.560$\pm$ 0.013   \\
$\omega(^\circ)$&77$^\circ.1 \pm 1^\circ.3$&74$^\circ.9 \pm 1^\circ.3$ 
&78$^\circ.5 \pm 1^\circ.5$\\
$ K_{1}$(\kms) &30.8 $\pm$ 0.3       &27.8 $\pm$ 0.6      &31.2 $\pm$ 0.6    \\
$ K_{2}$(\kms) &  ...                &66.2 $\pm$ 3.6      &62.2 $\pm$ 1.4    \\
$\gamma$(\kms) &-11.6$\pm$0.2        &-10.1$\pm$0.2       &-10.7$\pm$0.3     \\
$m_{A}$        &  ...                & 5.5 $\pm$ 0.5      & 3.50$\pm$ 0.20   \\
$m_{B}$        &  ...                & 2.3 $\pm$ 0.2      & 1.75$\pm$ 0.08   \\
\end{tabular}
\end{small}
\end{flushleft}
\end{table}
                                                                      
Our orbital solution yields a mass ratio 0.50 $\pm$ 0.03, which agrees with the
masses from the mass-luminosity relation.  Figure 1 compares the observed and
computed binary spectra in the 4957 \AA\, spectral region for a few orbital
phases in the rest frame of the primary.  The contributions from the secondary
are shown by dashed lines.

\begin{figure}[hbt]
\centerline{
\psfig{figure=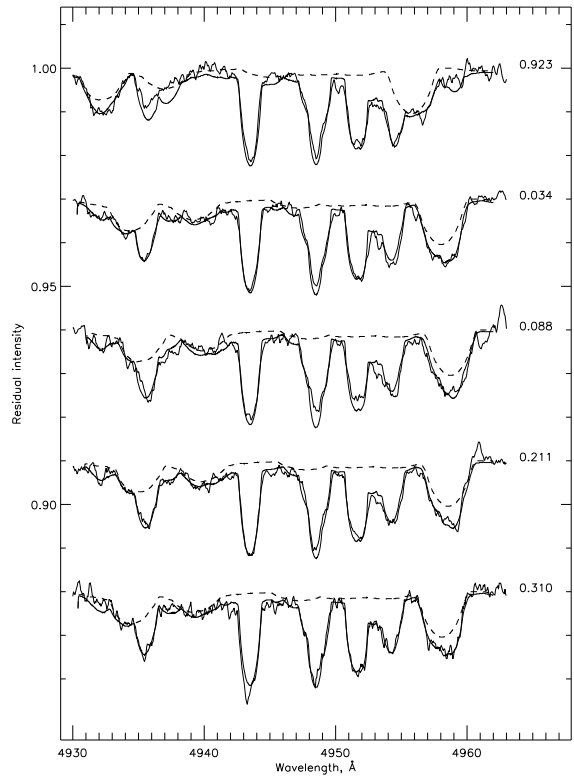,width=8cm}}
\caption[]{Comparison between the observed (thin lines) and the synthesized
(thick lines) spectra of \and\, in the $\lambda$ 4957 spectral region for
different orbital phases. The contributions from the secondary
are shown by the dashed lines.}
\label{4960_comp}
\end{figure}

\vspace{-2mm}
\section{Physical parameters of the \and\, components}

\vspace{-1mm}

The effective temperatures and the surface gravities of the components of \and\,
were obtained by fitting the observed spectrophotometry (Adelman \& Pyper 1983).
Combining data from the visual orbit measurements and from the spectroscopy and
using evolutionary tracks by Schaller et al. (1992) we found:

Primary: $\log(L/L_\odot)$ = 2.38 $\pm$ 0.14, $m=3.8 \pm 0.2 m_\odot$,
 $\rm R_{A}=2.7 \pm 0.4 \rm R_\odot$, $T_{\rm eff}$ = 13800 K, $\log\,
 {g}$ = 3.75, t=6$\cdot 10^7$
                                                   
Secondary: $\log(L/L_\odot)$ = 1.10 $\pm$ 0.2, $m=1.85 \pm 0.13 m_\odot$,
 $\rm R_{B}=1.65 \pm 0.3 \rm R_\odot$, $T_{\rm eff}$ = 8500 K, $\log\,
 {g}$ = 4.0, t=7$\cdot 10^7$

Both stars are close to the Zero Age Main Sequence.  The abundances of both
stars were derived by synthetic spectrum calculations (see Fig. 1). The chemical
abundances of \and\, A are typical of hot HgMn stars.  The secondary star shows
abundances typical of the metallic-line stars: a Ca deficiency, small
overabundances of the iron-peak~elements, and 1.0~dex overabundances of Sr
and Ba.

\acknowledgements
% Do not leave a blank line here! <---------------------->
This work has been partially supported by a Grant 1.4.1.5 of the Russian Federal
program ``Astronomy'' and by grants from The Citadel Development Foundation.

\end{document}